# The Optical Gravitational Lensing Experiment.
# The Catalog of Periodic Variable Stars in the Galactic Bulge.
# I. Periodic Variables in the Center of the Baade's Window*

by

A. U d a l s k i[1], M. K u b i a k[1], M. S z y m a ń s k i[1], J. K a ł u ż n y[1],
M. M a t e o[2] and W. K r z e m i ń s k i[3]

[1] Warsaw University Observatory, Al. Ujazdowskie 4, 00–478 Warszawa, Poland
e-mail: (udalski,mk,msz,jka)@sirius.astrouw.edu.pl

[2] Department of Astronomy, University of Michigan, 821 Dennison Bldg., Ann Arbor,
MI 48109–1090, USA
e-mail: mateo@astro.lsa.umich.edu

[3] Carnegie Observatories, Las Campanas Observatory, Casilla 601, La Serena, Chile
e-mail: wojtek@roses.ctio.noao.edu

### ABSTRACT

This paper is the first part of the Catalog of Periodic Variable Stars in the Galactic bulge. The Catalog is based on observations collected during the OGLE microlensing search. 213 periodic variable stars brighter than $I = 18^{\rm m}$: 31 pulsating, 116 eclipsing and 66 miscellaneous type variables from the Baade's Window BWC field are presented. Periodic variable stars from remaining 20 fields will be presented in similar form in the next parts of the Catalog. The Catalog as well as observations of all periodic variable objects are available to astronomical community over the Internet network.

## 1. Introduction

The Optical Gravitational Lensing Experiment (OGLE) is a long term observing project with the main goal of searching for dark matter in our Galaxy using microlensing (Paczyński 1986). After three seasons of continuous photometric monitoring of approximately two million non-variable stars in the direction of the Galactic bulge, twelve events have been detected (Udalski *et al.* 1993b, Udalski *et al.* 1994b, Udalski *et al.* 1994c).

---

The collected photometric data provide a unique material for studying the Galactic bulge (Paczyński *et al.* 1994, Stanek *et al.* 1994). Long span and large number of observations make it also possible to study precisely the population of variable stars toward the Galactic bulge with much deeper range and precision than searches attempted previously (*e.g.*, Blanco 1984). Analysis of variable stars in the Galactic bulge should shed a new light on the evolution of our Galaxy and evolutionary status of different class of variable stars. Eclipsing binaries will provide information allowing precise determination of the distance to the Galactic center. Also precise determination of the interstellar reddening toward the Galactic bulge should be possible.

This paper is the first, pilot paper, from a series of papers constituting the Catalog of Periodic Variable Stars in the Galactic bulge. The aim of the Catalog is to provide possibly most complete sample of periodic variable stars detected in selected areas of the Galactic bulge, namely those observed regularly during the OGLE microlensing search. We do not attempt detailed analysis of discovered variables here, as this will be a subject of following papers. The Catalog is thought to be an open publication with periodic updates when more data become available and/or additional search in sample of fainter stars is concluded. The photometric data for all detected variables are available to astronomical community over the Internet network.

In this paper we present 213 periodic variable stars detected in one field designated as BWC, that is the center of the Baade's Window. In the following papers the stars from remaining 11 Baade's Window fields will be presented, followed by data from 9 fields in other parts of the Galactic bulge. In Section 2 we describe shortly reduction and period search techniques. Section 3 presents the main rules of the Catalog. BWC field periodic variable stars are presented in Section 4 followed by short discussion of the completeness of the Catalog in Section 5.

## 2. Reductions and Period Search Technique

The Catalog of Periodic Variable Stars in the Galactic bulge is based on observations obtained during the OGLE microlensing search. Data presented here were collected during three observing OGLE seasons, 1992 – 1994, from April 19, 1992 up to September 16, 1994. Each observing season lasts approximately from April to September. Full logs of observations can be found in Udalski *et al.* (1992), Udalski *et al.* (1994a) and Udalski *et al.* (1995).

All observations were collected with the 1-m Swope telescope at Las Campanas Observatory which is operated by the Carnegie Institution of Washington. $2048 \times 2048$ Ford/Loral CCD with 15 $\mu$m pixels giving a scale of 0.44 arcsec/pixel was used as the detector. Due to strategy adopted for microlensing search the vast majority of observations was made in the $I$-band with sporadic $V$-band measurements.

The collected frames were fed into the standard OGLE data-pipeline, in which they were debiased and flat-fielded automatically and finally the photometry of objects was derived. Reductions of frames were done in the near real time. The modified DoPhot photometry program in the fixed position mode (Schechter *et al.* 1993) was used to derive profile-fitting photometry. Due to significant variations of the Point Spread Function (PSF) over the chip, the frame dividing method was developed: the frame was divided to 7 by 7 grid of slightly overlapping subframes where the PSF function could be assumed as constant. The photometry from all subframes was then tied to the common photometric system based on photometry on the overlapping parts of the subframes. Depending on quality of photometry a grade 'A' – 'F' was assigned to each frame with 'A' meaning the best and 'F' – unacceptable. Details of reduction procedure can be found in Udalski *et al.* (1992).

In order to manipulate efficiently the huge amount of data the databases for each observed field and band were created. Photometry of each frame with grade better than F was aperture corrected and included in the appropriate database. Details of database structure can be found in Szymański and Udalski (1993). Transformation to the standard *VI* system is described in Udalski *et al.* (1992). The error of the zero point of the photometry is not larger than 0.04 mag. The errors of individual observations depend on magnitude of the star and frame grade – detailed analysis of errors can be found in Udalski *et al.* (1992) and Udalski *et al.* (1994b).

21 fields – $15' \times 15'$ each – were observed frequently enough that variable stars search could be performed. In this paper we present periodic variable stars from the center of the Baade's Window field designated as BWC. The J2000 coordinates of the BWC field are: $\alpha = 18^{\mathrm{h}}03^{\mathrm{m}}24^{\mathrm{s}}$, $\delta = -30°02'00''$ ($l = 1°\!.0$ $b = -3°\!.9$).

Due to large number of observed stars we decided to limit the first edition of the Catalog to stars brighter than $I = 18.0$ mag. In the following updates the Catalog will be extended to fainter stars. The Catalog has also an upper limit of brightness due to saturation of bright star images on the CCD detector. That limit is equal to $I \approx 14^{\mathrm{m}}$. Thus some bright class of variables like Mira-type long period pulsating stars known to be present in the Galactic bulge are absent in the Catalog.

Stars searched for periodic variability were extracted from the $I$-band database in the following way. First, a star had to have at least 40 good quality measurements. The measurement was defined as "good" when its magnitude error was smaller than 1.6 of the median of magnitude error from the entire set of observation of the star (for stars with large amplitude like Algol eclipsing variables this factor was set to 3). Only 'A' – 'D' grade frames were included in the analysis. Second, the standard deviation of the magnitude had to be larger than the sigma limit for non-variable star of a given magnitude. The limiting curves "maximum sigma for non-variable stars *vs.* magnitude" were derived from the analysis of the distribution of standard deviations of all stars (Udalski *et al.* 1993a). From 38390 objects with good photometry and $I < 18^{\mathrm{m}}$ a total of 15759 stars suspected for variability were selected. Typically the number of good measurements of a suspected star was in the range of 100 – 170. Table 1 lists statistics of the BWC field stars searched for variability in one-magnitude bins.

All analyzed stars were subject to period search procedure. Analysis of variance method (AoV) (Schwarzenberg-Czerny 1989) of period determination was chosen as main period search technique. Tests showed that the AoV method was the most reliable method allowing selection of not only typical periodic variables but also objects like Algol eclipsing binaries often missed using other techniques. Compared with other methods, the AoV turned out to have the least number of misses. The

T a b l e 1

Number of stars searched for variability in one-magnitude bins

| $I$-magnitude | 14.5 | 15.5 | 16.5 | 17.5 |
|---|---|---|---|---|
| Total number of stars | 2200 | 5786 | 7560 | 22844 |
| Number of stars suspected for variability | 489 | 1675 | 3245 | 10350 |
| Periodic variable stars | 11 | 55 | 68 | 79 |

period search was limited to periods from 0.1 day to 100 days. The lower limit of AoV statistic value was chosen to be 10. This corresponds to approximately 0.005 significance level of the derived period. To minimize number of spurious periods around 0.5, 1.0, 1.5, 2.0, 2.5 and 3.0 days due to aliasing, only periods with AoV statistic greater than 25 were further analyzed in period ranges within 1% around those periods.

Observations of selected objects were then folded with the periods derived using AoV method, light curves were inspected visually and objects with random variability were removed from the list of periodic variables. The final list of variable stars included only those stars which exhibited unambiguous periodic variability.

To derive $V-I$ color of the variable star at maximum light, its $V$-band magnitudes were extracted from the $V$ filter database. In some cases $V$ magnitudes could not be retrieved when the star was not present in the $V$ database, typically due to bad photometry. Because of small number of $V$ measurements the $V$-band data were folded with the period derived from $I$-band data and $V$ magnitude at maximum was then interpolated. The accuracy of derived $V$-band maximum brightness magnitude is different for different objects but is usually better than 0.05 magnitude.

Equatorial coordinates of the variable stars were calculated using transformation "frame position – equatorial coordinates" derived using stars from the HST Guide Star Catalog (Lasker *et al.* 1988). 21 GSC stars were identified in the BWC field and used to determine transformation. The relative accuracy of derived coordinates is about $0\rlap{.}''2$ while the absolute accuracy is worse – about $1''$. The mean differences between coordinates derived for 25 RR Lyr stars from Blanco (1984) list (see Section 4.1) and their coordinates provided by Blanco are $\Delta\alpha = -0\rlap{.}''12 \pm 0\rlap{.}''79$ $\Delta\delta = 1\rlap{.}''58 \pm 2\rlap{.}''60$.

### 3. Catalog of Periodic Variable Stars

The Catalog of Periodic Variables in the Galactic bulge consists of two parts: a table which contains numerical data for all periodic objects and an atlas of light curves and finding charts. The variable star is designated according to the following scheme: OGLE *field* V*number*, where *field* is the general name of the OGLE field (see Table 1 in Udalski *et al.* 1994a) and *number* is a unique consecutive number of the variable star in a given field. Initially the variable stars list is sorted according to maximum magnitude. Thus, lower number means brighter star. Unfortunately this convention will be broken after the first update of the Catalog. However, as the majority of brighter stars has already been discovered, the general rule will be held with few exceptions. In the case when the variable star lies in the overlapping area between two fields, the star will appear in the Catalog only once with a designation of the field searched earlier.

The periodic variable stars are grouped in three categories: pulsating stars, eclipsing stars and miscellaneous stars. Assignment to one of the group is based on the shape of light curve and in some ambiguous cases on location of the star in the color-magnitude diagram (CMD). In some cases classification is doubtful and it is possible that a star is misclassified. Only spectral observations may solve the ambiguity. The group of miscellaneous variable stars is a "grab bag" of periodic stars which show sinusoidal-like variability and cannot be unambiguously classified as pulsating or eclipsing. Vast majority of stars in this group are most likely chromospherically active late type spotted stars or ellipsoidal variables. Classification within pulsating and eclipsing variable groups follows general scheme of General Catalog of Variable Stars (Kholopov *et al.* 1985).

For each group the table with numerical data contains the following information:

1. Star designation.

2. Right ascension (J2000).

3. Declination (J2000).

4. Period in days.

5. JD hel. of maximum brightness (minimum for eclipsing stars).

6. $I$ magnitude at maximum brightness.

7. $V - I$ at maximum brightness.

8. $I$ amplitude.

9. Classification.

10. Remarks.

Figures in the atlas show light curves and finding charts for all objects. The light curve is repeated twice for clarity of the figure. Phase zero corresponds to minimum of light for eclipsing stars (eclipse) and maximum for other type of stars. The finding chart is a $30'' \times 30''$ $I$-band subframe centered on the variable star. The variable star is marked by two horizontal and vertical bars. North is up and East is to the left. If the star is not located in the center of the finding chart it means that it is close to the edge of the whole field.

## 4. Catalog of Periodic Variable Stars of the BWC Field

### 4.1. Pulsating Stars

Table 2 and Appendix A present the catalog of pulsating objects from the BWC field. 31 variable stars were classified as pulsating stars. Most of them are RR Lyr

Table 2

Pulsating Variable Stars in the BWC field

| Star ID OGLE | $\alpha_{2000}$ | $\delta_{2000}$ | $P$ | $T_0 - 2\,448\,000$ | $I$ | $(V-I)$ | $\Delta I$ | Type | Remarks |
|---|---|---|---|---|---|---|---|---|---|
| BWC V1 | $18^h 03^m 33\overset{s}{.}60$ | $-30°01'14\overset{''}{.}4$ | $1\overset{d}{.}74795$ | 723.4328 | $13\overset{m}{.}92$ | $0\overset{m}{.}92$ | $0\overset{m}{.}79$ | ACEP | Blanco58 |
| BWC V6 | $18^h 03^m 16\overset{s}{.}08$ | $-30°01'40\overset{''}{.}2$ | $0\overset{d}{.}42765$ | 724.8069 | $14\overset{m}{.}77$ | $1\overset{m}{.}07$ | $0\overset{m}{.}53$ | RRab | Blanco46 |
| BWC V14 | $18^h 03^m 23\overset{s}{.}26$ | $-30°02'46\overset{''}{.}7$ | $0\overset{d}{.}44022$ | 724.3842 | $15\overset{m}{.}15$ | $0\overset{m}{.}66$ | $0\overset{m}{.}89$ | RRab | Blanco51 |
| BWC V15 | $18^h 03^m 24\overset{s}{.}15$ | $-30°05'12\overset{''}{.}6$ | $0\overset{d}{.}45871$ | 724.4866 | $15\overset{m}{.}16$ | $0\overset{m}{.}82$ | $0\overset{m}{.}83$ | RRab | Blanco52 |
| BWC V17 | $18^h 03^m 33\overset{s}{.}60$ | $-30°03'08\overset{''}{.}8$ | $0\overset{d}{.}29872$ | 724.8411 | $15\overset{m}{.}20$ | $0\overset{m}{.}87$ | $0\overset{m}{.}30$ | RRc | Blanco59 |
| BWC V22 | $18^h 03^m 39\overset{s}{.}74$ | $-30°08'59\overset{''}{.}5$ | $0\overset{d}{.}48968$ | 724.3396 | $15\overset{m}{.}25$ | $0\overset{m}{.}86$ | $0\overset{m}{.}85$ | RRab | Blanco62 |
| BWC V23 | $18^h 03^m 18\overset{s}{.}64$ | $-30°01'08\overset{''}{.}5$ | $0\overset{d}{.}45426$ | 724.8536 | $15\overset{m}{.}27$ | $0\overset{m}{.}70$ | $0\overset{m}{.}88$ | RRab | Blanco47 |
| BWC V25 | $18^h 03^m 35\overset{s}{.}20$ | $-30°01'24\overset{''}{.}3$ | $0\overset{d}{.}47418$ | 724.5423 | $15\overset{m}{.}32$ | $1\overset{m}{.}09$ | $0\overset{m}{.}72$ | RRab | Blanco60 |
| BWC V26 | $18^h 02^m 54\overset{s}{.}87$ | $-29°59'56\overset{''}{.}2$ | $0\overset{d}{.}47863$ | 724.5357 | $15\overset{m}{.}34$ | $1\overset{m}{.}35$ | $0\overset{m}{.}85$ | RRab | Blanco31 |
| BWC V28 | $18^h 03^m 51\overset{s}{.}71$ | $-30°04'48\overset{''}{.}2$ | $0\overset{d}{.}59478$ | 724.1030 | $15\overset{m}{.}40$ | $1\overset{m}{.}09$ | $0\overset{m}{.}46$ | RRab | Blanco65 |
| BWC V30 | $18^h 03^m 18\overset{s}{.}65$ | $-30°05'50\overset{''}{.}8$ | $0\overset{d}{.}57147$ | 724.4811 | $15\overset{m}{.}43$ | $1\overset{m}{.}06$ | $0\overset{m}{.}47$ | RRab | Blanco48 |
| BWC V33 | $18^h 02^m 57\overset{s}{.}32$ | $-30°00'05\overset{''}{.}3$ | $0\overset{d}{.}55032$ | 724.5403 | $15\overset{m}{.}46$ | $1\overset{m}{.}11$ | $0\overset{m}{.}62$ | RRab | Blanco32 |
| BWC V35 | $18^h 03^m 29\overset{s}{.}64$ | $-30°07'30\overset{''}{.}2$ | $0\overset{d}{.}33048$ | 724.6815 | $15\overset{m}{.}48$ | $1\overset{m}{.}00$ | $0\overset{m}{.}28$ | RRc | Blanco56 |
| BWC V37 | $18^h 02^m 49\overset{s}{.}00$ | $-30°00'51\overset{''}{.}3$ | $0\overset{d}{.}38016$ | 724.6135 | $15\overset{m}{.}56$ | $0\overset{m}{.}99$ | $0\overset{m}{.}32$ | RRc | Blanco29 |
| BWC V41 | $18^h 03^m 15\overset{s}{.}83$ | $-30°08'56\overset{''}{.}5$ | $0\overset{d}{.}46214$ | 724.4414 | $15\overset{m}{.}57$ | $1\overset{m}{.}32$ | $0\overset{m}{.}68$ | RRab | Blanco45 |
| BWC V47 | $18^h 03^m 29\overset{s}{.}86$ | $-30°01'56\overset{''}{.}0$ | $0\overset{d}{.}25692$ | 724.6249 | $15\overset{m}{.}64$ | $0\overset{m}{.}88$ | $0\overset{m}{.}26$ | RRc | Blanco54 |
| BWC V48 | $18^h 03^m 19\overset{s}{.}94$ | $-30°06'53\overset{''}{.}3$ | $0\overset{d}{.}33546$ | 724.8150 | $15\overset{m}{.}66$ | $1\overset{m}{.}08$ | $0\overset{m}{.}30$ | RRc | Blanco50 |
| BWC V51 | $18^h 03^m 17\overset{s}{.}10$ | $-29°57'23\overset{''}{.}2$ | $0\overset{d}{.}64949$ | 724.5158 | $15\overset{m}{.}69$ | $1\overset{m}{.}16$ | $0\overset{m}{.}24$ | RRab | |
| BWC V54 | $18^h 03^m 36\overset{s}{.}29$ | $-30°01'26\overset{''}{.}9$ | $0\overset{d}{.}28870$ | 724.6803 | $15\overset{m}{.}72$ | $0\overset{m}{.}93$ | $0\overset{m}{.}32$ | RRc | Blanco61 |
| BWC V56 | $18^h 02^m 51\overset{s}{.}17$ | $-30°09'54\overset{''}{.}4$ | $0\overset{d}{.}68046$ | 724.3327 | $15\overset{m}{.}72$ | $1\overset{m}{.}47$ | $0\overset{m}{.}20$ | RRab | |
| BWC V59 | $18^h 03^m 29\overset{s}{.}95$ | $-30°01'29\overset{''}{.}5$ | $0\overset{d}{.}26995$ | 724.6983 | $15\overset{m}{.}76$ | $0\overset{m}{.}85$ | $0\overset{m}{.}34$ | RRc | Blanco55 |
| BWC V60 | $18^h 03^m 11\overset{s}{.}80$ | $-30°02'45\overset{''}{.}2$ | $0\overset{d}{.}32069$ | 724.7402 | $15\overset{m}{.}78$ | $1\overset{m}{.}01$ | $0\overset{m}{.}27$ | RRc | Blanco44 |
| BWC V61 | $18^h 02^m 51\overset{s}{.}90$ | $-30°08'02\overset{''}{.}8$ | $0\overset{d}{.}61595$ | 724.3880 | $15\overset{m}{.}79$ | $1\overset{m}{.}29$ | $0\overset{m}{.}25$ | RRab | Blanco30 |
| BWC V62 | $18^h 03^m 40\overset{s}{.}22$ | $-30°02'47\overset{''}{.}1$ | $0\overset{d}{.}28682$ | 724.7640 | $15\overset{m}{.}80$ | $1\overset{m}{.}02$ | $0\overset{m}{.}30$ | RRc | Blanco63 |
| BWC V65 | $18^h 02^m 46\overset{s}{.}44$ | $-29°59'21\overset{''}{.}7$ | $0\overset{d}{.}55720$ | 724.7946 | $15\overset{m}{.}88$ | $0\overset{m}{.}49$ | $0\overset{m}{.}81$ | RRab | Blanco25 |
| BWC V81 | $18^h 03^m 06\overset{s}{.}86$ | $-29°56'55\overset{''}{.}2$ | $0\overset{d}{.}38590$ | 724.4462 | $16\overset{m}{.}23$ | $1\overset{m}{.}21$ | $0\overset{m}{.}23$ | RRab? | |
| BWC V82 | $18^h 03^m 42\overset{s}{.}57$ | $-30°01'34\overset{''}{.}1$ | $0\overset{d}{.}16127$ | 724.8409 | $16\overset{m}{.}25$ | $1\overset{m}{.}03$ | $0\overset{m}{.}30$ | SXPHE | Blanco64 |
| BWC V89 | $18^h 02^m 46\overset{s}{.}46$ | $-29°59'21\overset{''}{.}2$ | $0\overset{d}{.}55729$ | 724.1371 | $16\overset{m}{.}35$ | $0\overset{m}{.}09$ | $0\overset{m}{.}52$ | RRab? | |
| BWC V97 | $18^h 03^m 55\overset{s}{.}39$ | $-30°07'36\overset{''}{.}9$ | $0\overset{d}{.}11581$ | 724.9115 | $16\overset{m}{.}43$ | $1\overset{m}{.}23$ | $0\overset{m}{.}11$ | SXPHE | |
| BWC V106 | $18^h 03^m 03\overset{s}{.}73$ | $-30°03'59\overset{''}{.}8$ | $0\overset{d}{.}46496$ | 724.6262 | $16\overset{m}{.}66$ | $1\overset{m}{.}54$ | $0\overset{m}{.}56$ | RRab | Blanco37 |
| BWC V150 | $18^h 02^m 56\overset{s}{.}64$ | $-29°56'27\overset{''}{.}3$ | $0\overset{d}{.}07982$ | 724.9013 | $17\overset{m}{.}11$ | $0\overset{m}{.}99$ | $0\overset{m}{.}30$ | SXPHE | |

type stars of type ab and c. Three short period SX Phe type pulsating stars were also found. The period of the most prominent, OGLE BWC V150, falls beyond the limit of the period search, but the star was detected with the $2 \times P$. The remaining star in this group is an anomalous Cepheid (OGLE BWC V1).

Fig. 1 shows the CMD for the BWC field with pulsating stars denoted by open circles. Only 20% of BWC non-variable stars were plotted in the CMD for clarity. The distribution of stars in the CMD can be found in Paczyński *et al.* (1994). Most of pulsating stars are located on the horizontal branch of BWC. One of the RR Lyr stars: OGLE BWC V15 shows variable amplitude and period, *i.e.*, the Blazhko effect.

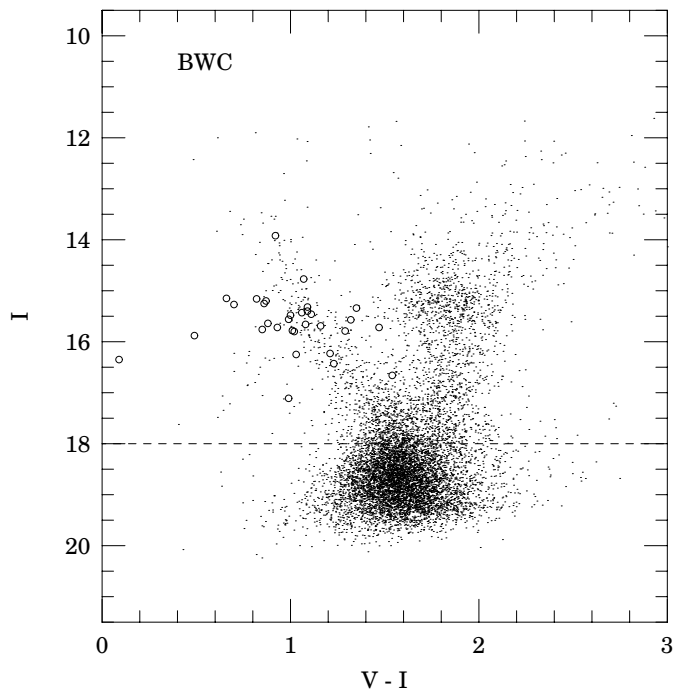

Fig. 1. CMD of the BWC field with positions of pulsating stars at maximum brightness denoted by open circles. Only 20% of non-variable stars were plotted for clarity. Dashed horizontal line marks limit of the present search for periodic variable stars.

The RR Lyr stars near the center of the Baade's Window were studied earlier by Blanco (1984). 25 objects from the Blanco list were identified in our data; cross-identification is given in "remarks" column. The agreement between our periods and those derived by Blanco is excellent in all but one (OGLE BWC V61) cases.

*4.2. Eclipsing Stars*

Table 3 and Appendix B contain the catalog of eclipsing variables from the BWC field. 116 variable stars of this type were discovered. The stars were

Table 3

Eclipsing Variable Stars in the BWC field

| Star ID OGLE | $\alpha_{2000}$ | $\delta_{2000}$ | $P$ | $T_0 -$ 2 448 000 | $I$ | $(V-I)$ | $\Delta I$ | Type | Remarks |
|---|---|---|---|---|---|---|---|---|---|
| BWC V4 | $18^h 03^m 06\overset{s}{.}64$ | $-30°05'23\overset{''}{.}5$ | $0\overset{d}{.}40260$ | 724.3619 | $14\overset{m}{.}63$ | $1\overset{m}{.}15$ | $0\overset{m}{.}19$ | EW | |
| BWC V18 | $18^h 02^m 56\overset{s}{.}96$ | $-30°07'30\overset{''}{.}3$ | $0\overset{d}{.}60114$ | 724.4939 | $15\overset{m}{.}21$ | $1\overset{m}{.}10$ | $0\overset{m}{.}73$ | EW | |
| BWC V19 | $18^h 03^m 48\overset{s}{.}47$ | $-29°58'07\overset{''}{.}3$ | $1\overset{d}{.}18572$ | 723.6851 | $15\overset{m}{.}21$ | $0\overset{m}{.}95$ | $1\overset{m}{.}07$ | EA | |
| BWC V20 | $18^h 02^m 47\overset{s}{.}03$ | $-30°02'26\overset{''}{.}1$ | $0\overset{d}{.}75862$ | 724.2849 | $15\overset{m}{.}24$ | $1\overset{m}{.}40$ | $0\overset{m}{.}17$ | EW | |
| BWC V24 | $18^h 03^m 46\overset{s}{.}09$ | $-30°01'46\overset{''}{.}9$ | $0\overset{d}{.}47158$ | 724.4577 | $15\overset{m}{.}30$ | $1\overset{m}{.}21$ | $0\overset{m}{.}20$ | EW | |
| BWC V36 | $18^h 03^m 01\overset{s}{.}49$ | $-30°08'43\overset{''}{.}8$ | $26\overset{d}{.}30714$ | 693.9824 | $15\overset{m}{.}51$ | $1\overset{m}{.}82$ | $0\overset{m}{.}24$ | EB | |
| BWC V43 | $18^h 03^m 46\overset{s}{.}34$ | $-29°56'45\overset{''}{.}1$ | $0\overset{d}{.}93020$ | 724.2723 | $15\overset{m}{.}58$ | $1\overset{m}{.}00$ | $0\overset{m}{.}31$ | EW | |
| BWC V50 | $18^h 03^m 34\overset{s}{.}05$ | $-29°56'47\overset{''}{.}0$ | $0\overset{d}{.}39730$ | 724.4527 | $15\overset{m}{.}68$ | $1\overset{m}{.}09$ | $0\overset{m}{.}16$ | EW | |
| BWC V53 | $18^h 03^m 06\overset{s}{.}46$ | $-29°56'27\overset{''}{.}6$ | $1\overset{d}{.}18892$ | 724.3792 | $15\overset{m}{.}69$ | $1\overset{m}{.}24$ | $0\overset{m}{.}13$ | EB | |
| BWC V57 | $18^h 03^m 14\overset{s}{.}95$ | $-30°08'07\overset{''}{.}6$ | $0\overset{d}{.}34142$ | 724.6981 | $15\overset{m}{.}74$ | $1\overset{m}{.}52$ | $0\overset{m}{.}27$ | EW | |
| BWC V58 | $18^h 03^m 18\overset{s}{.}76$ | $-30°04'00\overset{''}{.}6$ | $0\overset{d}{.}61188$ | 724.1936 | $15\overset{m}{.}75$ | $1\overset{m}{.}12$ | $0\overset{m}{.}23$ | EW | |
| BWC V64 | $18^h 03^m 47\overset{s}{.}07$ | $-30°03'53\overset{''}{.}0$ | $6\overset{d}{.}14613$ | 714.2110 | $15\overset{m}{.}82$ | $1\overset{m}{.}93$ | $0\overset{m}{.}31$ | EA | |
| BWC V66 | $18^h 02^m 49\overset{s}{.}95$ | $-29°58'37\overset{''}{.}7$ | $1\overset{d}{.}21492$ | 723.6732 | $15\overset{m}{.}98$ | $1\overset{m}{.}20$ | $0\overset{m}{.}11$ | EB | |
| BWC V68 | $18^h 03^m 40\overset{s}{.}08$ | $-30°07'20\overset{''}{.}4$ | $0\overset{d}{.}34860$ | 724.8222 | $16\overset{m}{.}02$ | $1\overset{m}{.}55$ | $0\overset{m}{.}16$ | EW | |
| BWC V70 | $18^h 03^m 33\overset{s}{.}26$ | $-29°58'48\overset{''}{.}3$ | $1\overset{d}{.}23965$ | 723.5691 | $16\overset{m}{.}04$ | $1\overset{m}{.}14$ | $0\overset{m}{.}60$ | EA | |
| BWC V73 | $18^h 02^m 51\overset{s}{.}72$ | $-30°05'03\overset{''}{.}1$ | $2\overset{d}{.}76618$ | 722.6092 | $16\overset{m}{.}08$ | $1\overset{m}{.}53$ | $0\overset{m}{.}25$ | EA | |
| BWC V75 | $18^h 03^m 19\overset{s}{.}55$ | $-30°00'32\overset{''}{.}2$ | $1\overset{d}{.}06538$ | 724.0193 | $16\overset{m}{.}10$ | $1\overset{m}{.}84$ | $0\overset{m}{.}87$ | EA | |
| BWC V76 | $18^h 03^m 02\overset{s}{.}16$ | $-30°05'07\overset{''}{.}3$ | $0\overset{d}{.}37768$ | 724.5507 | $16\overset{m}{.}11$ | $2\overset{m}{.}22$ | $0\overset{m}{.}16$ | EW | |
| BWC V79 | $18^h 03^m 08\overset{s}{.}45$ | $-30°08'46\overset{''}{.}1$ | $1\overset{d}{.}19855$ | 722.9294 | $16\overset{m}{.}19$ | $1\overset{m}{.}09$ | $0\overset{m}{.}14$ | EA | |
| BWC V80 | $18^h 02^m 53\overset{s}{.}42$ | $-29°59'30\overset{''}{.}6$ | $0\overset{d}{.}70392$ | 724.1753 | $16\overset{m}{.}18$ | $1\overset{m}{.}87$ | $0\overset{m}{.}12$ | EW | |
| BWC V85 | $18^h 03^m 20\overset{s}{.}08$ | $-30°02'17\overset{''}{.}1$ | $1\overset{d}{.}52482$ | 724.1828 | $16\overset{m}{.}30$ | $1\overset{m}{.}03$ | $0\overset{m}{.}34$ | EA | |
| BWC V86 | $18^h 03^m 24\overset{s}{.}31$ | $-29°55'16\overset{''}{.}7$ | $2\overset{d}{.}32321$ | 724.3420 | $16\overset{m}{.}29$ | $1\overset{m}{.}24$ | $0\overset{m}{.}14$ | E | |
| BWC V88 | $18^h 03^m 22\overset{s}{.}58$ | $-30°06'25\overset{''}{.}5$ | $0\overset{d}{.}41456$ | 724.3192 | $16\overset{m}{.}30$ | $1\overset{m}{.}34$ | $0\overset{m}{.}36$ | EW | |
| BWC V90 | $18^h 03^m 01\overset{s}{.}09$ | $-30°09'08\overset{''}{.}3$ | $0\overset{d}{.}94896$ | 723.6721 | $16\overset{m}{.}35$ | $1\overset{m}{.}29$ | $0\overset{m}{.}42$ | EW | |
| BWC V91 | $18^h 03^m 14\overset{s}{.}15$ | $-30°06'09\overset{''}{.}8$ | $3\overset{d}{.}36931$ | 724.3343 | $16\overset{m}{.}43$ | – | $0\overset{m}{.}20$ | EA | |
| BWC V92 | $18^h 03^m 00\overset{s}{.}36$ | $-30°00'24\overset{''}{.}3$ | $0\overset{d}{.}46736$ | 724.4244 | $16\overset{m}{.}42$ | $1\overset{m}{.}42$ | $0\overset{m}{.}35$ | EW | |
| BWC V93 | $18^h 03^m 11\overset{s}{.}72$ | $-30°05'48\overset{''}{.}3$ | $0\overset{d}{.}41138$ | 724.4525 | $16\overset{m}{.}42$ | $1\overset{m}{.}39$ | $0\overset{m}{.}25$ | EW | |
| BWC V95 | $18^h 03^m 40\overset{s}{.}60$ | $-30°04'24\overset{''}{.}6$ | $4\overset{d}{.}79478$ | 721.1663 | $16\overset{m}{.}43$ | $1\overset{m}{.}22$ | $1\overset{m}{.}62$ | EA | |
| BWC V96 | $18^h 03^m 04\overset{s}{.}96$ | $-30°07'11\overset{''}{.}6$ | $1\overset{d}{.}22146$ | 723.7639 | $16\overset{m}{.}43$ | $1\overset{m}{.}37$ | $0\overset{m}{.}32$ | EB | |
| BWC V98 | $18^h 03^m 36\overset{s}{.}04$ | $-30°07'12\overset{''}{.}4$ | $0\overset{d}{.}45720$ | 724.4105 | $16\overset{m}{.}44$ | $1\overset{m}{.}36$ | $0\overset{m}{.}13$ | EW | |
| BWC V100 | $18^h 02^m 55\overset{s}{.}20$ | $-30°07'37\overset{''}{.}5$ | $3\overset{d}{.}14845$ | 721.9553 | $16\overset{m}{.}45$ | $1\overset{m}{.}31$ | $0\overset{m}{.}23$ | EA | |
| BWC V101 | $18^h 03^m 28\overset{s}{.}75$ | $-30°07'31\overset{''}{.}2$ | $0\overset{d}{.}55078$ | 724.6447 | $16\overset{m}{.}49$ | $1\overset{m}{.}28$ | $0\overset{m}{.}52$ | EW | |
| BWC V102 | $18^h 03^m 04\overset{s}{.}85$ | $-29°58'03\overset{''}{.}2$ | $0\overset{d}{.}28812$ | 724.8984 | $16\overset{m}{.}52$ | $1\overset{m}{.}69$ | $0\overset{m}{.}42$ | EW | ? |
| BWC V103 | $18^h 02^m 57\overset{s}{.}88$ | $-29°55'05\overset{''}{.}4$ | $3\overset{d}{.}47268$ | 718.9142 | $16\overset{m}{.}57$ | – | $0\overset{m}{.}85$ | EA | |
| BWC V104 | $18^h 03^m 20\overset{s}{.}21$ | $-30°03'14\overset{''}{.}9$ | $0\overset{d}{.}79237$ | 723.9211 | $16\overset{m}{.}56$ | $1\overset{m}{.}29$ | $0\overset{m}{.}46$ | EA | |
| BWC V107 | $18^h 03^m 29\overset{s}{.}51$ | $-30°02'26\overset{''}{.}1$ | $1\overset{d}{.}94564$ | 723.1458 | $16\overset{m}{.}68$ | $1\overset{m}{.}66$ | $0\overset{m}{.}35$ | EA | |
| BWC V109 | $18^h 03^m 24\overset{s}{.}38$ | $-30°07'08\overset{''}{.}4$ | $0\overset{d}{.}28902$ | 724.8506 | $16\overset{m}{.}71$ | $1\overset{m}{.}73$ | $0\overset{m}{.}24$ | EW | |
| BWC V110 | $18^h 03^m 51\overset{s}{.}22$ | $-30°03'14\overset{''}{.}5$ | $0\overset{d}{.}41648$ | 724.4420 | $16\overset{m}{.}71$ | $1\overset{m}{.}45$ | $0\overset{m}{.}48$ | EW | |
| BWC V111 | $18^h 03^m 11\overset{s}{.}15$ | $-29°57'06\overset{''}{.}1$ | $0\overset{d}{.}63908$ | 724.2153 | $16\overset{m}{.}74$ | $1\overset{m}{.}42$ | $0\overset{m}{.}75$ | EW | |
| BWC V112 | $18^h 03^m 00\overset{s}{.}44$ | $-30°08'30\overset{''}{.}3$ | $0\overset{d}{.}73334$ | 724.2356 | $16\overset{m}{.}75$ | $1\overset{m}{.}33$ | $0\overset{m}{.}12$ | EW | |
| BWC V115 | $18^h 03^m 40\overset{s}{.}76$ | $-30°07'55\overset{''}{.}6$ | $0\overset{d}{.}30248$ | 724.5092 | $16\overset{m}{.}79$ | $1\overset{m}{.}51$ | $0\overset{m}{.}29$ | EW | |
| BWC V116 | $18^h 03^m 31\overset{s}{.}79$ | $-30°03'43\overset{''}{.}1$ | $0\overset{d}{.}57444$ | 724.3832 | $16\overset{m}{.}80$ | $1\overset{m}{.}25$ | $0\overset{m}{.}32$ | EW | |
| BWC V117 | $18^h 03^m 06\overset{s}{.}00$ | $-29°57'32\overset{''}{.}7$ | $2\overset{d}{.}42807$ | 722.8243 | $16\overset{m}{.}83$ | $2\overset{m}{.}05$ | $0\overset{m}{.}46$ | EA | |
| BWC V118 | $18^h 03^m 31\overset{s}{.}17$ | $-29°54'42\overset{''}{.}4$ | $0\overset{d}{.}25152$ | 725.1787 | $16\overset{m}{.}85$ | $1\overset{m}{.}20$ | $0\overset{m}{.}14$ | EW | |
| BWC V120 | $18^h 03^m 36\overset{s}{.}09$ | $-29°55'24\overset{''}{.}1$ | $1\overset{d}{.}33062$ | 722.9192 | $16\overset{m}{.}85$ | $1\overset{m}{.}18$ | $1\overset{m}{.}13$ | EA | |
| BWC V121 | $18^h 03^m 23\overset{s}{.}41$ | $-29°58'00\overset{''}{.}3$ | $0\overset{d}{.}32440$ | 724.6350 | $16\overset{m}{.}87$ | $1\overset{m}{.}57$ | $0\overset{m}{.}21$ | EW | |
| BWC V123 | $18^h 03^m 05\overset{s}{.}40$ | $-29°58'49\overset{''}{.}2$ | $0\overset{d}{.}66901$ | 724.0528 | $16\overset{m}{.}90$ | $1\overset{m}{.}46$ | $0\overset{m}{.}32$ | EW | |
| BWC V124 | $18^h 03^m 09\overset{s}{.}46$ | $-30°07'16\overset{''}{.}2$ | $0\overset{d}{.}68898$ | 724.3281 | $16\overset{m}{.}92$ | $1\overset{m}{.}34$ | $0\overset{m}{.}35$ | EW | |
| BWC V125 | $18^h 03^m 46\overset{s}{.}80$ | $-30°01'22\overset{''}{.}2$ | $0\overset{d}{.}44070$ | 724.5989 | $16\overset{m}{.}92$ | $1\overset{m}{.}51$ | $0\overset{m}{.}19$ | EW | |
| BWC V126 | $18^h 03^m 24\overset{s}{.}54$ | $-29°56'36\overset{''}{.}2$ | $1\overset{d}{.}67149$ | 723.4543 | $16\overset{m}{.}94$ | $1\overset{m}{.}10$ | $1\overset{m}{.}03$ | EA | |
| BWC V127 | $18^h 02^m 52\overset{s}{.}17$ | $-30°09'09\overset{''}{.}7$ | $0\overset{d}{.}75417$ | 724.1012 | $16\overset{m}{.}94$ | $1\overset{m}{.}43$ | $0\overset{m}{.}17$ | EW | |

T a b l e 3

Continued

| Star ID OGLE | $\alpha_{2000}$ | $\delta_{2000}$ | $P$ | $T_0 -$ 2 448 000 | $I$ | $(V-I)$ | $\Delta I$ | Type | Remarks |
|---|---|---|---|---|---|---|---|---|---|
| BWC V128 | $18^h02^m58\overset{s}{.}88$ | $-30°05'57\overset{''}{.}5$ | $0\overset{d}{.}38330$ | 724.7551 | $16\overset{m}{.}94$ | $1\overset{m}{.}53$ | $0\overset{m}{.}43$ | EW | |
| BWC V129 | $18^h02^m49\overset{s}{.}21$ | $-30°03'10\overset{''}{.}1$ | $0\overset{d}{.}27362$ | 724.7291 | $16\overset{m}{.}96$ | $1\overset{m}{.}53$ | $0\overset{m}{.}88$ | EW | |
| BWC V131 | $18^h03^m08\overset{s}{.}27$ | $-30°01'39\overset{''}{.}2$ | $0\overset{d}{.}60336$ | 724.6082 | $16\overset{m}{.}97$ | $1\overset{m}{.}34$ | $0\overset{m}{.}21$ | EW | |
| BWC V136 | $18^h03^m20\overset{s}{.}47$ | $-30°00'18\overset{''}{.}4$ | $0\overset{d}{.}76094$ | 723.8213 | $17\overset{m}{.}02$ | $1\overset{m}{.}29$ | $0\overset{m}{.}22$ | EW | |
| BWC V137 | $18^h03^m44\overset{s}{.}76$ | $-29°58'20\overset{''}{.}7$ | $1\overset{d}{.}94931$ | 722.8431 | $17\overset{m}{.}05$ | – | $0\overset{m}{.}36$ | E | |
| BWC V138 | $18^h02^m49\overset{s}{.}17$ | $-30°05'39\overset{''}{.}7$ | $0\overset{d}{.}51920$ | 724.6838 | $17\overset{m}{.}03$ | $1\overset{m}{.}96$ | $0\overset{m}{.}25$ | EW | |
| BWC V139 | $18^h03^m20\overset{s}{.}97$ | $-30°09'01\overset{''}{.}0$ | $0\overset{d}{.}56038$ | 724.5293 | $17\overset{m}{.}04$ | $1\overset{m}{.}47$ | $0\overset{m}{.}52$ | E | |
| BWC V141 | $18^h03^m41\overset{s}{.}98$ | $-30°07'44\overset{''}{.}1$ | $3\overset{d}{.}81546$ | 723.0866 | $17\overset{m}{.}04$ | $1\overset{m}{.}37$ | $0\overset{m}{.}53$ | EA | |
| BWC V142 | $18^h02^m53\overset{s}{.}12$ | $-30°06'13\overset{''}{.}4$ | $0\overset{d}{.}54052$ | 724.7346 | $17\overset{m}{.}05$ | $1\overset{m}{.}37$ | $0\overset{m}{.}28$ | EW | |
| BWC V143 | $18^h02^m57\overset{s}{.}10$ | $-29°59'54\overset{''}{.}6$ | $1\overset{d}{.}15833$ | 722.9253 | $17\overset{m}{.}06$ | $1\overset{m}{.}22$ | $0\overset{m}{.}79$ | EA | ? |
| BWC V145 | $18^h02^m59\overset{s}{.}50$ | $-30°00'50\overset{''}{.}6$ | $0\overset{d}{.}53352$ | 724.2321 | $17\overset{m}{.}08$ | $1\overset{m}{.}58$ | $0\overset{m}{.}32$ | EW | |
| BWC V146 | $18^h03^m21\overset{s}{.}26$ | $-30°09'20\overset{''}{.}1$ | $0\overset{d}{.}59236$ | 724.3260 | $17\overset{m}{.}08$ | $1\overset{m}{.}37$ | $0\overset{m}{.}35$ | EW | |
| BWC V147 | $18^h03^m34\overset{s}{.}77$ | $-30°03'45\overset{''}{.}9$ | $0\overset{d}{.}59642$ | 724.0688 | $17\overset{m}{.}09$ | $1\overset{m}{.}15$ | $0\overset{m}{.}33$ | EW | |
| BWC V148 | $18^h03^m19\overset{s}{.}02$ | $-30°01'10\overset{''}{.}0$ | $0\overset{d}{.}44290$ | 724.4427 | $17\overset{m}{.}09$ | $1\overset{m}{.}31$ | $0\overset{m}{.}20$ | EW | |
| BWC V149 | $18^h02^m57\overset{s}{.}01$ | $-30°05'56\overset{''}{.}0$ | $3\overset{d}{.}50343$ | 720.7957 | $17\overset{m}{.}10$ | $1\overset{m}{.}45$ | $0\overset{m}{.}21$ | E | |
| BWC V151 | $18^h03^m19\overset{s}{.}16$ | $-30°01'13\overset{''}{.}3$ | $1\overset{d}{.}47328$ | 724.6619 | $17\overset{m}{.}12$ | $1\overset{m}{.}11$ | $0\overset{m}{.}48$ | EA | |
| BWC V153 | $18^h03^m47\overset{s}{.}68$ | $-30°05'26\overset{''}{.}0$ | $0\overset{d}{.}74912$ | 724.1144 | $17\overset{m}{.}13$ | $1\overset{m}{.}37$ | $0\overset{m}{.}47$ | EW | |
| BWC V154 | $18^h03^m37\overset{s}{.}41$ | $-29°58'07\overset{''}{.}0$ | $0\overset{d}{.}40602$ | 724.4209 | $17\overset{m}{.}20$ | – | $0\overset{m}{.}46$ | EW | |
| BWC V155 | $18^h02^m49\overset{s}{.}04$ | $-30°05'44\overset{''}{.}2$ | $0\overset{d}{.}47466$ | 724.5115 | $17\overset{m}{.}20$ | $1\overset{m}{.}58$ | $0\overset{m}{.}25$ | EW | |
| BWC V156 | $18^h03^m30\overset{s}{.}52$ | $-29°55'12\overset{''}{.}7$ | $1\overset{d}{.}84553$ | 722.7465 | $17\overset{m}{.}21$ | $1\overset{m}{.}37$ | $0\overset{m}{.}53$ | EA | |
| BWC V157 | $18^h03^m48\overset{s}{.}40$ | $-30°01'58\overset{''}{.}6$ | $0\overset{d}{.}58614$ | 724.1437 | $17\overset{m}{.}21$ | $1\overset{m}{.}32$ | $0\overset{m}{.}33$ | EW | |
| BWC V158 | $18^h03^m03\overset{s}{.}06$ | $-30°03'41\overset{''}{.}2$ | $0\overset{d}{.}56256$ | 724.3860 | $17\overset{m}{.}22$ | $1\overset{m}{.}50$ | $0\overset{m}{.}31$ | EW | |
| BWC V159 | $18^h03^m30\overset{s}{.}36$ | $-30°07'57\overset{''}{.}8$ | $0\overset{d}{.}86158$ | 724.2355 | $17\overset{m}{.}22$ | $1\overset{m}{.}47$ | $0\overset{m}{.}70$ | EA | |
| BWC V160 | $18^h03^m55\overset{s}{.}07$ | $-30°07'43\overset{''}{.}9$ | $0\overset{d}{.}91660$ | 724.0957 | $17\overset{m}{.}21$ | $2\overset{m}{.}05$ | $0\overset{m}{.}21$ | EW | |
| BWC V161 | $18^h02^m52\overset{s}{.}14$ | $-30°00'35\overset{''}{.}4$ | $0\overset{d}{.}98974$ | 724.3115 | $17\overset{m}{.}24$ | $1\overset{m}{.}74$ | $0\overset{m}{.}14$ | EW | |
| BWC V163 | $18^h03^m22\overset{s}{.}81$ | $-29°55'27\overset{''}{.}6$ | $0\overset{d}{.}48918$ | 724.7887 | $17\overset{m}{.}25$ | $1\overset{m}{.}39$ | $0\overset{m}{.}25$ | EW | |
| BWC V165 | $18^h03^m27\overset{s}{.}60$ | $-30°02'11\overset{''}{.}1$ | $0\overset{d}{.}49918$ | 724.4649 | $17\overset{m}{.}28$ | $1\overset{m}{.}43$ | $0\overset{m}{.}33$ | EW | |
| BWC V166 | $18^h03^m32\overset{s}{.}91$ | $-29°56'41\overset{''}{.}1$ | $0\overset{d}{.}49758$ | 724.6568 | $17\overset{m}{.}29$ | $1\overset{m}{.}56$ | $0\overset{m}{.}44$ | EW | |
| BWC V168 | $18^h03^m10\overset{s}{.}07$ | $-30°00'02\overset{''}{.}8$ | $0\overset{d}{.}58226$ | 724.5556 | $17\overset{m}{.}32$ | $1\overset{m}{.}65$ | $0\overset{m}{.}20$ | EW | |
| BWC V169 | $18^h03^m06\overset{s}{.}86$ | $-29°56'06\overset{''}{.}7$ | $0\overset{d}{.}35265$ | 724.5247 | $17\overset{m}{.}35$ | $1\overset{m}{.}89$ | $0\overset{m}{.}20$ | EW | |
| BWC V170 | $18^h03^m20\overset{s}{.}46$ | $-29°59'19\overset{''}{.}9$ | $0\overset{d}{.}37944$ | 724.4895 | $17\overset{m}{.}36$ | $1\overset{m}{.}27$ | $0\overset{m}{.}20$ | EW | |
| BWC V171 | $18^h03^m06\overset{s}{.}16$ | $-30°00'56\overset{''}{.}8$ | $0\overset{d}{.}53260$ | 724.8340 | $17\overset{m}{.}41$ | $1\overset{m}{.}60$ | $0\overset{m}{.}27$ | EW | |
| BWC V172 | $18^h03^m37\overset{s}{.}83$ | $-29°56'33\overset{''}{.}8$ | $0\overset{d}{.}45128$ | 724.6338 | $17\overset{m}{.}41$ | $1\overset{m}{.}40$ | $0\overset{m}{.}51$ | EW | |
| BWC V174 | $18^h03^m39\overset{s}{.}53$ | $-30°06'50\overset{''}{.}4$ | $0\overset{d}{.}45654$ | 724.9125 | $17\overset{m}{.}43$ | $1\overset{m}{.}66$ | $0\overset{m}{.}46$ | EW | |
| BWC V175 | $18^h03^m44\overset{s}{.}57$ | $-29°57'02\overset{''}{.}2$ | $0\overset{d}{.}52104$ | 724.3029 | $17\overset{m}{.}43$ | $1\overset{m}{.}37$ | $0\overset{m}{.}24$ | EW | |
| BWC V176 | $18^h02^m51\overset{s}{.}04$ | $-30°07'33\overset{''}{.}6$ | $0\overset{d}{.}70219$ | 723.9836 | $17\overset{m}{.}44$ | $1\overset{m}{.}62$ | $0\overset{m}{.}43$ | EA | |
| BWC V177 | $18^h03^m12\overset{s}{.}14$ | $-29°59'53\overset{''}{.}1$ | $0\overset{d}{.}47792$ | 724.3810 | $17\overset{m}{.}44$ | $1\overset{m}{.}52$ | $0\overset{m}{.}24$ | EW | |
| BWC V178 | $18^h03^m26\overset{s}{.}26$ | $-30°03'38\overset{''}{.}2$ | $0\overset{d}{.}51894$ | 724.9540 | $17\overset{m}{.}44$ | $1\overset{m}{.}39$ | $0\overset{m}{.}44$ | EW | |
| BWC V179 | $18^h03^m08\overset{s}{.}15$ | $-29°57'06\overset{''}{.}5$ | $0\overset{d}{.}45808$ | 724.3459 | $17\overset{m}{.}45$ | $1\overset{m}{.}57$ | $0\overset{m}{.}34$ | EW | |
| BWC V181 | $18^h02^m54\overset{s}{.}49$ | $-29°57'21\overset{''}{.}2$ | $0\overset{d}{.}46218$ | 724.2823 | $17\overset{m}{.}47$ | $1\overset{m}{.}69$ | $0\overset{m}{.}43$ | EW | |
| BWC V182 | $18^h03^m27\overset{s}{.}75$ | $-30°02'46\overset{''}{.}9$ | $0\overset{d}{.}52430$ | 724.4673 | $17\overset{m}{.}47$ | $1\overset{m}{.}43$ | $0\overset{m}{.}55$ | EW | |
| BWC V183 | $18^h03^m32\overset{s}{.}00$ | $-30°05'44\overset{''}{.}4$ | $3\overset{d}{.}05953$ | 721.6819 | $17\overset{m}{.}52$ | – | $0\overset{m}{.}36$ | EA | ? |
| BWC V186 | $18^h03^m07\overset{s}{.}90$ | $-30°00'21\overset{''}{.}6$ | $0\overset{d}{.}37394$ | 724.4340 | $17\overset{m}{.}52$ | $1\overset{m}{.}31$ | $0\overset{m}{.}32$ | EW | |
| BWC V187 | $18^h03^m50\overset{s}{.}90$ | $-29°57'10\overset{''}{.}8$ | $0\overset{d}{.}50304$ | 724.4204 | $17\overset{m}{.}52$ | $1\overset{m}{.}75$ | $0\overset{m}{.}53$ | EW | |
| BWC V188 | $18^h03^m14\overset{s}{.}92$ | $-30°05'38\overset{''}{.}6$ | $0\overset{d}{.}40366$ | 724.7574 | $17\overset{m}{.}54$ | $1\overset{m}{.}29$ | $0\overset{m}{.}43$ | EW | |
| BWC V189 | $18^h03^m17\overset{s}{.}86$ | $-30°02'41\overset{''}{.}6$ | $0\overset{d}{.}40486$ | 724.5091 | $17\overset{m}{.}53$ | $1\overset{m}{.}84$ | $0\overset{m}{.}65$ | EW | |
| BWC V190 | $18^h03^m39\overset{s}{.}88$ | $-29°55'42\overset{''}{.}3$ | $1\overset{d}{.}08715$ | 723.0986 | $17\overset{m}{.}57$ | $1\overset{m}{.}35$ | $0\overset{m}{.}43$ | EA | |
| BWC V191 | $18^h03^m38\overset{s}{.}77$ | $-29°56'44\overset{''}{.}0$ | $0\overset{d}{.}38572$ | 724.5519 | $17\overset{m}{.}59$ | $1\overset{m}{.}56$ | $0\overset{m}{.}34$ | EW | |
| BWC V192 | $18^h03^m22\overset{s}{.}96$ | $-30°03'50\overset{''}{.}7$ | $0\overset{d}{.}52618$ | 724.3262 | $17\overset{m}{.}63$ | – | $0\overset{m}{.}48$ | EW | |
| BWC V194 | $18^h03^m47\overset{s}{.}62$ | $-30°01'30\overset{''}{.}8$ | $1\overset{d}{.}62209$ | 722.4140 | $17\overset{m}{.}60$ | $1\overset{m}{.}68$ | $0\overset{m}{.}75$ | EA | |
| BWC V195 | $18^h02^m53\overset{s}{.}16$ | $-30°07'09\overset{''}{.}9$ | $0\overset{d}{.}39624$ | 724.5526 | $17\overset{m}{.}62$ | $1\overset{m}{.}57$ | $0\overset{m}{.}27$ | EW | |

T a b l e 3

Concluded

| Star ID OGLE | $\alpha_{2000}$ | $\delta_{2000}$ | $P$ | $T_0 -$ 2 448 000 | $I$ | $(V-I)$ | $\Delta I$ | Type | Remarks |
|---|---|---|---|---|---|---|---|---|---|
| BWC V197 | $18^h 03^m 45\overset{s}{.}54$ | $-29°56'13\overset{''}{.}7$ | $0\overset{d}{.}37862$ | 724.6446 | $17\overset{m}{.}64$ | $1\overset{m}{.}44$ | $0\overset{m}{.}35$ | EW | |
| BWC V198 | $18^h 02^m 45\overset{s}{.}79$ | $-29°58'34\overset{''}{.}8$ | $0\overset{d}{.}72938$ | 723.9060 | $17\overset{m}{.}66$ | $1\overset{m}{.}18$ | $0\overset{m}{.}22$ | EW | |
| BWC V199 | $18^h 03^m 27\overset{s}{.}70$ | $-30°04'57\overset{''}{.}0$ | $1\overset{d}{.}24217$ | 723.3591 | $17\overset{m}{.}65$ | $1\overset{m}{.}52$ | $1\overset{m}{.}14$ | EA | |
| BWC V201 | $18^h 02^m 49\overset{s}{.}98$ | $-30°01'48\overset{''}{.}1$ | $0\overset{d}{.}37432$ | 724.6423 | $17\overset{m}{.}68$ | $1\overset{m}{.}59$ | $0\overset{m}{.}28$ | EW | |
| BWC V202 | $18^h 03^m 14\overset{s}{.}45$ | $-30°05'55\overset{''}{.}6$ | $0\overset{d}{.}41214$ | 724.6452 | $17\overset{m}{.}68$ | $1\overset{m}{.}47$ | $0\overset{m}{.}24$ | EW | |
| BWC V203 | $18^h 03^m 44\overset{s}{.}61$ | $-30°01'54\overset{''}{.}7$ | $2\overset{d}{.}70091$ | 721.8961 | $17\overset{m}{.}68$ | $1\overset{m}{.}88$ | $0\overset{m}{.}67$ | EA | |
| BWC V204 | $18^h 02^m 48\overset{s}{.}84$ | $-29°57'35\overset{''}{.}0$ | $0\overset{d}{.}58100$ | 724.0887 | $17\overset{m}{.}70$ | $1\overset{m}{.}49$ | $0\overset{m}{.}77$ | EW | |
| BWC V205 | $18^h 03^m 47\overset{s}{.}40$ | $-30°06'12\overset{''}{.}9$ | $0\overset{d}{.}42100$ | 724.5494 | $17\overset{m}{.}70$ | $1\overset{m}{.}60$ | $0\overset{m}{.}37$ | EW | |
| BWC V206 | $18^h 03^m 35\overset{s}{.}52$ | $-30°03'49\overset{''}{.}0$ | $1\overset{d}{.}67204$ | 723.7843 | $17\overset{m}{.}75$ | $1\overset{m}{.}59$ | $0\overset{m}{.}71$ | EA | |
| BWC V207 | $18^h 03^m 20\overset{s}{.}22$ | $-30°00'00\overset{''}{.}4$ | $0\overset{d}{.}32904$ | 724.6053 | $17\overset{m}{.}76$ | $1\overset{m}{.}59$ | $0\overset{m}{.}38$ | EW | |
| BWC V208 | $18^h 03^m 32\overset{s}{.}01$ | $-30°07'32\overset{''}{.}1$ | $0\overset{d}{.}50680$ | 724.4478 | $17\overset{m}{.}77$ | $1\overset{m}{.}43$ | $0\overset{m}{.}56$ | EW | |
| BWC V210 | $18^h 02^m 57\overset{s}{.}25$ | $-30°00'25\overset{''}{.}1$ | $1\overset{d}{.}95712$ | 723.9093 | $17\overset{m}{.}80$ | $2\overset{m}{.}12$ | $0\overset{m}{.}67$ | EA | ? |
| BWC V211 | $18^h 03^m 23\overset{s}{.}15$ | $-30°01'42\overset{''}{.}0$ | $1\overset{d}{.}23911$ | 724.7431 | $17\overset{m}{.}81$ | $1\overset{m}{.}46$ | $0\overset{m}{.}54$ | EA | |
| BWC V213 | $18^h 03^m 14\overset{s}{.}35$ | $-29°58'24\overset{''}{.}9$ | $0\overset{d}{.}97215$ | 723.7316 | $17\overset{m}{.}85$ | $1\overset{m}{.}28$ | $0\overset{m}{.}32$ | E | |

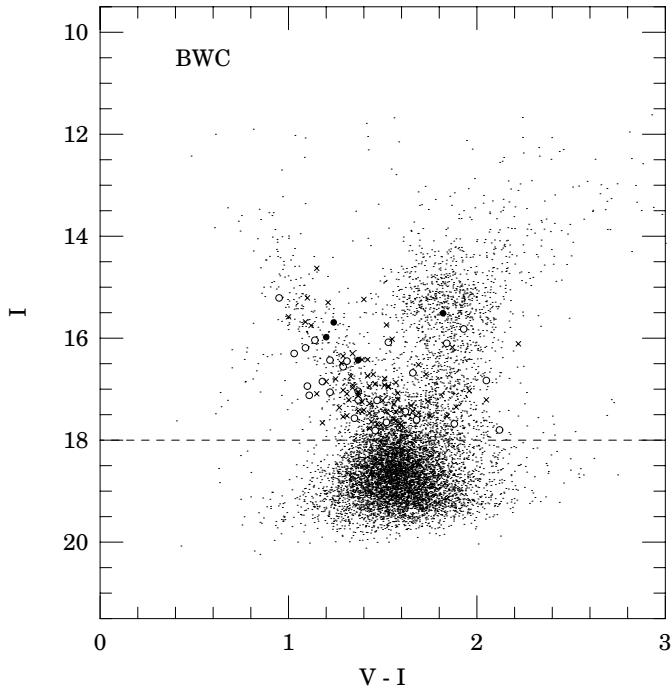

Fig. 2. Same as Fig. 1. Open circles denote Algol type eclipsing variables, filled circles – $\beta$ Lyr type stars and crosses – W-UMa type stars.

classified as Algol type – EA (30), $\beta$ Lyr type – EB (4), and W UMa type – EW (77). In 5 ambiguous cases only E type was assigned. For Algol type variables a total of at least five points distributed on both branches of the primary eclipse were required.

Fig. 2 shows the CMD for the BWC field with Algols (open circles), $\beta$ Lyr (filled circles), and W UMa type stars (crosses). Most of eclipsing stars is located on the narrow main sequence of disk stars (Paczyński *et al.* 1994).

### *4.3. Miscellaneous Stars*

Table 4 and Appendix C contain the catalog for miscellaneous group of periodic variable stars found in the BWC field. 66 stars were assigned to this group of variables with sinusoidal shape light curves. Majority of them are stars with periods longer than a few days. They are usually red subgiants and giants – most likely spotted, chromospherically active stars. In a few cases (*e.g.*, OGLE BWC V10, OGLE BWC V16, OGLE BWC V34) the amplitude of variations changes significantly from season to season whereas the period is essentially constant. This strongly supports chromospheric activity as the most plausible explanation of variability of these objects. Similar objects have already been observed (*e.g.*, Strassmeier *et al.* 1993), however, their large number is worth noting.

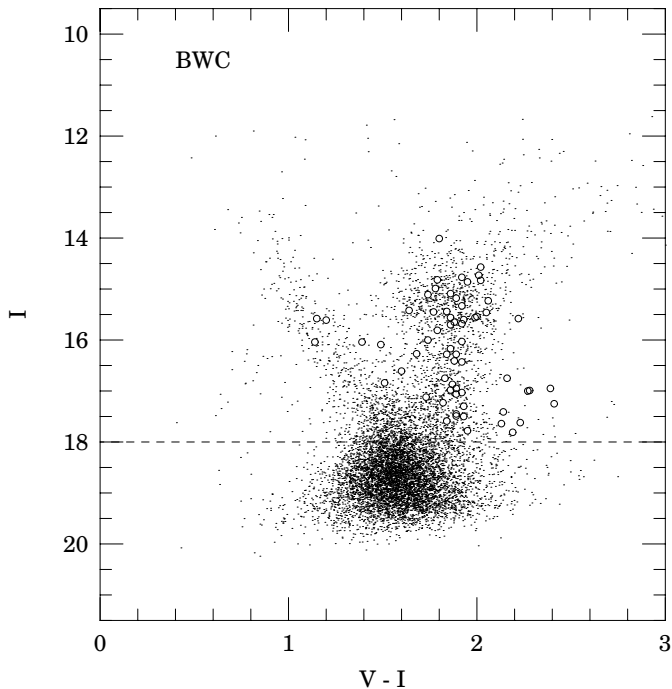

Fig. 3. Same as Fig. 1. Open circles show position of periodic variable stars classified as miscellaneous type.

Some of the objects (*e.g.*, OGLE BWC V39, OGLE BWC V72) might be ellipsoidal variables (these stars are marked "Ell" in remarks column of Table 4); if so, the true period should be twice of that listed in Table 4.

Fig. 3 presents the CMD with miscellaneous group stars shown as open circles.

T a b l e 4

Miscellaneous group of periodic variable stars in the BWC field

| Star ID OGLE | $\alpha_{2000}$ | $\delta_{2000}$ | $P$ | $T_0 -$ 2 448 000 | $I$ | $(V-I)$ | $\Delta I$ | Type | Remarks |
|---|---|---|---|---|---|---|---|---|---|
| BWC V2 | $18^h 03^m 46\overset{s}{.}41$ | $-30°04'17\overset{''}{.}0$ | $32\overset{d}{.}97609$ | 661.1706 | $14\overset{m}{.}01$ | $1\overset{m}{.}80$ | $0\overset{m}{.}12$ | MISC | |
| BWC V3 | $18^h 03^m 35\overset{s}{.}22$ | $-29°55'32\overset{''}{.}4$ | $16\overset{d}{.}71542$ | 704.8893 | $14\overset{m}{.}57$ | $2\overset{m}{.}02$ | $0\overset{m}{.}12$ | MISC | |
| BWC V5 | $18^h 03^m 26\overset{s}{.}26$ | $-30°05'40\overset{''}{.}9$ | $55\overset{d}{.}86592$ | 655.8659 | $14\overset{m}{.}73$ | $2\overset{m}{.}01$ | $0\overset{m}{.}19$ | MISC | |
| BWC V7 | $18^h 03^m 18\overset{s}{.}50$ | $-29°57'39\overset{''}{.}4$ | $49\overset{d}{.}44376$ | 646.9716 | $14\overset{m}{.}77$ | $1\overset{m}{.}92$ | $0\overset{m}{.}20$ | MISC | |
| BWC V8 | $18^h 03^m 50\overset{s}{.}51$ | $-29°57'16\overset{''}{.}0$ | $45\overset{d}{.}45455$ | 675.4546 | $14\overset{m}{.}82$ | $1\overset{m}{.}79$ | $0\overset{m}{.}10$ | MISC | |
| BWC V9 | $18^h 03^m 05\overset{s}{.}80$ | $-30°08'30\overset{''}{.}8$ | $16\overset{d}{.}24036$ | 704.6692 | $14\overset{m}{.}83$ | $2\overset{m}{.}02$ | $0\overset{m}{.}08$ | MISC | |
| BWC V10 | $18^h 03^m 06\overset{s}{.}53$ | $-29°56'38\overset{''}{.}6$ | $37\overset{d}{.}73585$ | 688.6793 | $14\overset{m}{.}86$ | $1\overset{m}{.}95$ | $0\overset{m}{.}20$ | MISC | |
| BWC V11 | $18^h 03^m 52\overset{s}{.}12$ | $-30°03'22\overset{''}{.}3$ | $48\overset{d}{.}48485$ | 630.7879 | $14\overset{m}{.}99$ | $1\overset{m}{.}78$ | $0\overset{m}{.}14$ | MISC | |
| BWC V12 | $18^h 03^m 44\overset{s}{.}07$ | $-30°06'42\overset{''}{.}8$ | $23\overset{d}{.}86635$ | 695.7041 | $15\overset{m}{.}09$ | $1\overset{m}{.}86$ | $0\overset{m}{.}09$ | MISC | |
| BWC V13 | $18^h 03^m 33\overset{s}{.}94$ | $-30°07'24\overset{''}{.}8$ | $50\overset{d}{.}82592$ | 703.9390 | $15\overset{m}{.}11$ | $1\overset{m}{.}74$ | $0\overset{m}{.}10$ | MISC | |
| BWC V16 | $18^h 03^m 22\overset{s}{.}68$ | $-30°05'50\overset{''}{.}5$ | $26\overset{d}{.}42008$ | 710.7001 | $15\overset{m}{.}18$ | $1\overset{m}{.}89$ | $0\overset{m}{.}15$ | MISC | |
| BWC V21 | $18^h 03^m 09\overset{s}{.}37$ | $-30°05'45\overset{''}{.}9$ | $17\overset{d}{.}05030$ | 703.8364 | $15\overset{m}{.}23$ | $2\overset{m}{.}06$ | $0\overset{m}{.}19$ | MISC | |
| BWC V27 | $18^h 03^m 53\overset{s}{.}32$ | $-30°08'15\overset{''}{.}7$ | $33\overset{d}{.}11258$ | 662.5827 | $15\overset{m}{.}33$ | $1\overset{m}{.}92$ | $0\overset{m}{.}13$ | MISC | |
| BWC V29 | $18^h 03^m 10\overset{s}{.}69$ | $-29°55'33\overset{''}{.}2$ | $5\overset{d}{.}26870$ | 718.4399 | $15\overset{m}{.}42$ | $1\overset{m}{.}64$ | $0\overset{m}{.}21$ | MISC | |
| BWC V31 | $18^h 02^m 53\overset{s}{.}51$ | $-30°05'19\overset{''}{.}8$ | $22\overset{d}{.}33389$ | 709.7710 | $15\overset{m}{.}45$ | $1\overset{m}{.}77$ | $0\overset{m}{.}11$ | MISC | |
| BWC V32 | $18^h 02^m 47\overset{s}{.}19$ | $-30°00'56\overset{''}{.}6$ | $20\overset{d}{.}55498$ | 703.3914 | $15\overset{m}{.}44$ | $1\overset{m}{.}84$ | $0\overset{m}{.}24$ | MISC | |
| BWC V34 | $18^h 02^m 48\overset{s}{.}55$ | $-29°58'33\overset{''}{.}3$ | $40\overset{d}{.}48583$ | 682.3887 | $15\overset{m}{.}46$ | $2\overset{m}{.}05$ | $0\overset{m}{.}49$ | MISC | |
| BWC V38 | $18^h 03^m 53\overset{s}{.}87$ | $-30°05'58\overset{''}{.}8$ | $39\overset{d}{.}40887$ | 674.2858 | $15\overset{m}{.}54$ | $2\overset{m}{.}00$ | $0\overset{m}{.}26$ | MISC | |
| BWC V39 | $18^h 02^m 52\overset{s}{.}77$ | $-30°05'00\overset{''}{.}4$ | $5\overset{d}{.}82920$ | 718.3906 | $15\overset{m}{.}58$ | $1\overset{m}{.}15$ | $0\overset{m}{.}10$ | MISC | Ell |
| BWC V40 | $18^h 02^m 46\overset{s}{.}39$ | $-30°00'01\overset{''}{.}7$ | $8\overset{d}{.}29876$ | 714.1913 | $15\overset{m}{.}56$ | $1\overset{m}{.}99$ | $0\overset{m}{.}17$ | MISC | |
| BWC V42 | $18^h 03^m 38\overset{s}{.}60$ | $-30°08'08\overset{''}{.}8$ | $40\overset{d}{.}00000$ | 719.2000 | $15\overset{m}{.}56$ | $1\overset{m}{.}86$ | $0\overset{m}{.}22$ | MISC | |
| BWC V44 | $18^h 03^m 35\overset{s}{.}10$ | $-30°07'30\overset{''}{.}8$ | $62\overset{d}{.}89309$ | 630.8177 | $15\overset{m}{.}58$ | $2\overset{m}{.}22$ | $0\overset{m}{.}33$ | MISC | |
| BWC V45 | $18^h 03^m 21\overset{s}{.}56$ | $-30°02'29\overset{''}{.}1$ | $6\overset{d}{.}67334$ | 718.7854 | $15\overset{m}{.}61$ | $1\overset{m}{.}20$ | $0\overset{m}{.}10$ | MISC | Ell ? |
| BWC V46 | $18^h 02^m 54\overset{s}{.}66$ | $-30°00'38\overset{''}{.}2$ | $24\overset{d}{.}18380$ | 695.5261 | $15\overset{m}{.}60$ | $1\overset{m}{.}93$ | $0\overset{m}{.}16$ | MISC | |
| BWC V49 | $18^h 03^m 14\overset{s}{.}19$ | $-29°55'25\overset{''}{.}6$ | $21\overset{d}{.}32196$ | 714.0724 | $15\overset{m}{.}65$ | $1\overset{m}{.}88$ | $0\overset{m}{.}18$ | MISC | |
| BWC V52 | $18^h 03^m 08\overset{s}{.}21$ | $-29°59'58\overset{''}{.}9$ | $20\overset{d}{.}69322$ | 714.7438 | $15\overset{m}{.}68$ | $1\overset{m}{.}92$ | $0\overset{m}{.}10$ | MISC | |
| BWC V55 | $18^h 03^m 48\overset{s}{.}34$ | $-30°01'52\overset{''}{.}1$ | $3\overset{d}{.}41472$ | 724.2882 | $15\overset{m}{.}70$ | $1\overset{m}{.}86$ | $0\overset{m}{.}11$ | MISC | Ell ? |
| BWC V63 | $18^h 03^m 33\overset{s}{.}21$ | $-29°56'25\overset{''}{.}3$ | $28\overset{d}{.}77393$ | 710.0930 | $15\overset{m}{.}81$ | $1\overset{m}{.}79$ | $0\overset{m}{.}12$ | MISC | |
| BWC V67 | $18^h 02^m 51\overset{s}{.}37$ | $-29°59'03\overset{''}{.}1$ | $39\overset{d}{.}52569$ | 676.6798 | $16\overset{m}{.}00$ | $1\overset{m}{.}74$ | $0\overset{m}{.}13$ | MISC | |
| BWC V69 | $18^h 03^m 04\overset{s}{.}64$ | $-30°02'32\overset{''}{.}3$ | $3\overset{d}{.}48129$ | 724.0039 | $16\overset{m}{.}04$ | $1\overset{m}{.}14$ | $0\overset{m}{.}12$ | MISC | Ell ? |
| BWC V71 | $18^h 02^m 58\overset{s}{.}11$ | $-30°07'54\overset{''}{.}2$ | $5\overset{d}{.}23766$ | 718.0308 | $16\overset{m}{.}03$ | $1\overset{m}{.}92$ | $0\overset{m}{.}09$ | MISC | Ell ? |
| BWC V72 | $18^h 03^m 50\overset{s}{.}29$ | $-30°04'42\overset{''}{.}4$ | $5\overset{d}{.}31773$ | 720.1270 | $16\overset{m}{.}04$ | $1\overset{m}{.}39$ | $0\overset{m}{.}10$ | MISC | Ell ? |
| BWC V74 | $18^h 02^m 57\overset{s}{.}53$ | $-29°57'02\overset{''}{.}3$ | $45\overset{d}{.}81902$ | 646.5064 | $16\overset{m}{.}09$ | $1\overset{m}{.}49$ | $0\overset{m}{.}15$ | MISC | |
| BWC V77 | $18^h 02^m 57\overset{s}{.}60$ | $-30°02'42\overset{''}{.}6$ | $40\overset{d}{.}48583$ | 686.6397 | $16\overset{m}{.}20$ | – | $0\overset{m}{.}20$ | MISC | |
| BWC V78 | $18^h 02^m 54\overset{s}{.}53$ | $-30°00'04\overset{''}{.}0$ | $3\overset{d}{.}79291$ | 721.3736 | $16\overset{m}{.}17$ | $1\overset{m}{.}86$ | $0\overset{m}{.}13$ | MISC | |
| BWC V83 | $18^h 02^m 47\overset{s}{.}68$ | $-29°59'15\overset{''}{.}9$ | $55\overset{d}{.}63283$ | 673.1572 | $16\overset{m}{.}27$ | $1\overset{m}{.}68$ | $0\overset{m}{.}14$ | MISC | |
| BWC V84 | $18^h 03^m 04\overset{s}{.}31$ | $-30°04'56\overset{''}{.}8$ | $61\overset{d}{.}53846$ | 636.9231 | $16\overset{m}{.}28$ | $1\overset{m}{.}84$ | $0\overset{m}{.}11$ | MISC | |
| BWC V87 | $18^h 03^m 44\overset{s}{.}42$ | $-29°54'43\overset{''}{.}9$ | $14\overset{d}{.}55075$ | 698.4360 | $16\overset{m}{.}28$ | $1\overset{m}{.}89$ | $0\overset{m}{.}22$ | MISC | |
| BWC V94 | $18^h 03^m 43\overset{s}{.}94$ | $-30°07'38\overset{''}{.}6$ | $3\overset{d}{.}81643$ | 720.5038 | $16\overset{m}{.}41$ | $1\overset{m}{.}88$ | $0\overset{m}{.}12$ | MISC | Ell ? |
| BWC V99 | $18^h 03^m 34\overset{s}{.}13$ | $-30°03'06\overset{''}{.}2$ | $54\overset{d}{.}92400$ | 659.1452 | $16\overset{m}{.}43$ | $1\overset{m}{.}92$ | $0\overset{m}{.}27$ | MISC | |
| BWC V105 | $18^h 03^m 21\overset{s}{.}43$ | $-29°56'38\overset{''}{.}7$ | $2\overset{d}{.}31415$ | 724.0513 | $16\overset{m}{.}61$ | $1\overset{m}{.}60$ | $0\overset{m}{.}22$ | MISC | Ell ? |
| BWC V108 | $18^h 03^m 09\overset{s}{.}39$ | $-30°06'49\overset{''}{.}8$ | $5\overset{d}{.}54835$ | 720.8971 | $16\overset{m}{.}72$ | – | $0\overset{m}{.}17$ | MISC | Ell ? |
| BWC V113 | $18^h 03^m 30\overset{s}{.}92$ | $-29°55'32\overset{''}{.}8$ | $19\overset{d}{.}75309$ | 705.1853 | $16\overset{m}{.}75$ | $1\overset{m}{.}83$ | $0\overset{m}{.}28$ | MISC | |
| BWC V114 | $18^h 03^m 34\overset{s}{.}30$ | $-29°59'40\overset{''}{.}9$ | $15\overset{d}{.}37870$ | 712.4952 | $16\overset{m}{.}75$ | $2\overset{m}{.}16$ | $0\overset{m}{.}27$ | MISC | |
| BWC V119 | $18^h 03^m 38\overset{s}{.}04$ | $-29°59'37\overset{''}{.}5$ | $4\overset{d}{.}09207$ | 721.2683 | $16\overset{m}{.}84$ | $1\overset{m}{.}51$ | $0\overset{m}{.}20$ | MISC | Ell ? |
| BWC V122 | $18^h 03^m 51\overset{s}{.}84$ | $-30°06'24\overset{''}{.}5$ | $5\overset{d}{.}52792$ | 720.7855 | $16\overset{m}{.}87$ | $1\overset{m}{.}87$ | $0\overset{m}{.}20$ | MISC | Ell ? |
| BWC V130 | $18^h 03^m 43\overset{s}{.}52$ | $-30°02'54\overset{''}{.}0$ | $5\overset{d}{.}79123$ | 719.3866 | $16\overset{m}{.}95$ | $1\overset{m}{.}89$ | $0\overset{m}{.}20$ | MISC | |
| BWC V132 | $18^h 03^m 23\overset{s}{.}03$ | $-30°07'44\overset{''}{.}9$ | $2\overset{d}{.}11372$ | 720.7785 | $16\overset{m}{.}95$ | $2\overset{m}{.}39$ | $0\overset{m}{.}22$ | MISC | Ell ? |
| BWC V133 | $18^h 02^m 48\overset{s}{.}54$ | $-30°00'25\overset{''}{.}7$ | $15\overset{d}{.}67085$ | 711.3620 | $16\overset{m}{.}98$ | $1\overset{m}{.}86$ | $0\overset{m}{.}12$ | MISC | |
| BWC V134 | $18^h 03^m 24\overset{s}{.}66$ | $-30°07'19\overset{''}{.}5$ | $11\overset{d}{.}31862$ | 717.0346 | $16\overset{m}{.}99$ | $2\overset{m}{.}28$ | $0\overset{m}{.}18$ | MISC | |

Table 4

Concluded.

| Star ID OGLE | $\alpha_{2000}$ | $\delta_{2000}$ | $P$ | $T_0 -$ 2 448 000 | $I$ | $(V-I)$ | $\Delta I$ | Type | Remarks |
|---|---|---|---|---|---|---|---|---|---|
| BWC V135 | $18^h03^m39\overset{s}{.}62$ | $-30°01'57\overset{''}{.}3$ | $17\overset{d}{.}05030$ | 713.2141 | $17\overset{m}{.}00$ | $2\overset{m}{.}27$ | $0\overset{m}{.}18$ | MISC | |
| BWC V140 | $18^h03^m16\overset{s}{.}92$ | $-30°04'24\overset{''}{.}0$ | $67\overset{d}{.}22689$ | 653.4454 | $17\overset{m}{.}03$ | $1\overset{m}{.}92$ | $0\overset{m}{.}18$ | MISC | |
| BWC V144 | $18^h02^m59\overset{s}{.}67$ | $-30°01'49\overset{''}{.}1$ | $4\overset{d}{.}49792$ | 721.1065 | $17\overset{m}{.}06$ | $1\overset{m}{.}89$ | $0\overset{m}{.}20$ | MISC | |
| BWC V152 | $18^h03^m32\overset{s}{.}79$ | $-30°08'11\overset{''}{.}2$ | $22\overset{d}{.}94894$ | 707.4011 | $17\overset{m}{.}12$ | $1\overset{m}{.}73$ | $0\overset{m}{.}17$ | MISC | |
| BWC V162 | $18^h03^m00\overset{s}{.}26$ | $-30°05'44\overset{''}{.}0$ | $57\overset{d}{.}80347$ | 640.1734 | $17\overset{m}{.}23$ | $1\overset{m}{.}82$ | $0\overset{m}{.}25$ | MISC | |
| BWC V164 | $18^h03^m21\overset{s}{.}63$ | $-30°07'11\overset{''}{.}6$ | $7\overset{d}{.}62922$ | 717.2993 | $17\overset{m}{.}25$ | $2\overset{m}{.}41$ | $0\overset{m}{.}40$ | MISC | |
| BWC V167 | $18^h02^m57\overset{s}{.}30$ | $-30°06'26\overset{''}{.}9$ | $23\overset{d}{.}46041$ | 686.0997 | $17\overset{m}{.}30$ | $1\overset{m}{.}93$ | $0\overset{m}{.}22$ | MISC | |
| BWC V173 | $18^h02^m48\overset{s}{.}39$ | $-29°55'30\overset{''}{.}8$ | $12\overset{d}{.}80410$ | 707.6826 | $17\overset{m}{.}41$ | $2\overset{m}{.}14$ | $0\overset{m}{.}25$ | MISC | |
| BWC V180 | $18^h03^m17\overset{s}{.}12$ | $-30°05'18\overset{''}{.}6$ | $20\overset{d}{.}44990$ | 696.3191 | $17\overset{m}{.}45$ | $1\overset{m}{.}89$ | $0\overset{m}{.}31$ | MISC | |
| BWC V184 | $18^h02^m53\overset{s}{.}27$ | $-29°59'17\overset{''}{.}5$ | $6\overset{d}{.}31014$ | 717.5891 | $17\overset{m}{.}49$ | $1\overset{m}{.}89$ | $0\overset{m}{.}27$ | MISC | |
| BWC V185 | $18^h03^m46\overset{s}{.}87$ | $-30°01'29\overset{''}{.}6$ | $25\overset{d}{.}99090$ | 676.5431 | $17\overset{m}{.}50$ | $1\overset{m}{.}93$ | $0\overset{m}{.}26$ | MISC | |
| BWC V193 | $18^h03^m35\overset{s}{.}02$ | $-29°56'14\overset{''}{.}2$ | $13\overset{d}{.}35113$ | 722.5632 | $17\overset{m}{.}59$ | $1\overset{m}{.}84$ | $0\overset{m}{.}35$ | MISC | |
| BWC V196 | $18^h03^m32\overset{s}{.}89$ | $-30°08'52\overset{''}{.}0$ | $8\overset{d}{.}24402$ | 711.7062 | $17\overset{m}{.}62$ | $2\overset{m}{.}23$ | $0\overset{m}{.}27$ | MISC | |
| BWC V200 | $18^h03^m40\overset{s}{.}17$ | $-30°05'32\overset{''}{.}5$ | $11\overset{d}{.}36364$ | 709.7730 | $17\overset{m}{.}64$ | $2\overset{m}{.}13$ | $0\overset{m}{.}22$ | MISC | |
| BWC V209 | $18^h03^m19\overset{s}{.}81$ | $-30°07'15\overset{''}{.}0$ | $6\overset{d}{.}57030$ | 720.8933 | $17\overset{m}{.}78$ | $1\overset{m}{.}95$ | $0\overset{m}{.}36$ | MISC | |
| BWC V212 | $18^h03^m31\overset{s}{.}68$ | $-30°06'17\overset{''}{.}7$ | $6\overset{d}{.}59848$ | 716.3310 | $17\overset{m}{.}81$ | $2\overset{m}{.}19$ | $0\overset{m}{.}33$ | MISC | |

## 5. Completeness of the Catalog

A huge number of variable stars included in the Catalog will make it possible to study statistically population of different group of variable stars. However, it is crucial to have at least some information concerning completeness of the Catalog. The detailed statistical analysis of the completeness of the Catalog will be published elsewhere. Here, only crude estimate is presented.

Two factors determine completeness of the Catalog. First factor describes what fraction of all stars underwent period search. OGLE reduction procedure neglects stars which have bad photometry due to CCD defects, severe blending or are overexposed *etc.* on the so called "template" frame of a given field (Udalski *et al.* 1992). Such objects are marked and are not measured on other frames. Thus some fraction of objects is *a priori* missed. Also some objects which do not meet criteria of good photometry (Szymański and Udalski 1993) are flagged in the database. These objects were not analyzed for variability.

It is relatively easy to determine this factor of completeness by adding randomly artificial stars to the frame and checking how many of them have been recovered using standard reduction procedure. Such tests have already been performed for analysis of CMDs of the Galactic bulge (Udalski *et al.* 1993a). The recovery efficiency for stars brighter than $I \approx 18^m$ has been found to be relatively flat and equal to $\approx 80\%$ for the densest BWC field and $\approx 87\%$ for the least dense field TP8.

Independent estimate of this factor was obtained by comparing list of RR Lyr type variables from Blanco (1984) with those detected in the OGLE search. RR Lyr are bright, large amplitude variables and therefore other factors describing efficiency of detection should be practically one for this group. 27 stars from the Blanco (1984) list are located in the BWC field. 25 of them were found in the Catalog. Two lacking variables are indeed marked as bad objects in the BWC field database. Taking into account that the sample is small this seems to be consistent with the artificial stars test.

The second factor describes efficiency of the period search method itself. Our tests showed that the AoV method used for period search gives the best results for different types of periodic variables. However, it also overlooks some objects, in particular low amplitude, noisy variables. The factor of efficiency of period search is a function of amplitude, typical error of observations of a given object, and period. Its detailed analysis will be performed by numerical simulations. At present, we can only assess from a very limited sample of stars reduced twice on overlapping subframes of the BWC field that for amplitudes larger than 0.2 mag and stars brighter than $I \approx 17^{\mathrm{m}}$ the AoV efficiency factor is close to one and drops for fainter, more noisy objects.

One more independent test of the completeness of the Catalog was carried out. We compared the Catalog variable stars list with an independent search for variables from the BWC field performed using the phase dispersion minimization (PDM) method (Stellingwerf 1979). The PDM search used data from 1992 and 1993 seasons only. 96 stars from the PDM search were positively identified among 213 variables in the Catalog. Only 7 stars were not found in the Catalog. Three of them turned out to be flagged in the database and as such they were not searched for variability because their photometry might be erroneous. The four remaining objects turned out to be a small amplitude ($\approx 0.1$ mag), sinusoidal shape, usually noisy and faint variables evidently overlooked in the Catalog. We did not add any of these objects to the Catalog to keep it consistent with our algorithm of search.

### 6. Summary

We present here the first part of the Catalog of Periodic Variable Stars in the Galactic bulge – periodic variables discovered in the BWC field. 213 periodic objects brighter than $I = 18^{\mathrm{m}}$ were detected in this field. Catalog of periodic variable objects from 20 other fields observed during the OGLE microlensing search will be published in similar form in the forthcoming papers. The Catalog will be updated regularly when our search is extended to fainter stars, more data are available or additional fields are observed.

It should be stressed that some of the stars in the Catalog, particularly those with longer periods, might be misclassified or have badly derived periods (*e.g.*, due to 1-year aliasing). We expect a feedback from astronomers to correct any errors in the next editions of the Catalog.

The Catalog is available to astronomical community in electronic form over the Internet network using anonymous ftp service from *sirius.astrouw.edu.pl* host (148.81.8.1), directory */ogle/var_catalog*. See README file in this directory for details. Beside the Catalog, also $I$-band observations of all periodic objects are available. The data for each star contain JD hel. date of the observation, $I$-band magnitude and error of the observation.

**Acknowledgements.** It is a great pleasure to thank B. Paczyński for valuable suggestions and discussions. This project was supported with the Polish KBN grant 2P30400306 to A. Udalski and the NSF grants AST 9216494 to B. Paczyński, AST 9216830 to G.W. Preston